\documentclass[9pt,twocolumn]{extarticle}
\pdfoutput=1
\usepackage{soul}
\usepackage{titlesec} 

\usepackage[superscript, nomove]{cite}

\usepackage{float}
\usepackage{caption}
\usepackage{lipsum,ulem}
\usepackage[verbose]{placeins}

\usepackage{dsfont}
\usepackage{graphicx}
\usepackage{subcaption}
\usepackage[margin=0.9in]{geometry}
\usepackage[usenames,dvipsnames]{xcolor}
\usepackage[colorlinks,linkcolor=Blue,urlcolor=Blue,citecolor=Blue]{hyperref}
\usepackage{amsmath,amssymb}
\usepackage[squaren]{SIunits}

\usepackage{mathpazo}
\usepackage{courier}
\normalfont
\usepackage[T1]{fontenc}
\usepackage{caption}
\usepackage{upgreek}

\newcommand{\ket}[1]{\ensuremath{|#1\rangle\mkern-1mu}}

\newcommand{\ad}[1]{\textsuperscript{#1}\kern-2pt}

\makeatletter
\def\blx@maxline{77}
\makeatother

\usepackage{capt-of}

\def\mytitle{Probing the fractional quantum Hall phases in valley-layer locked bilayer MoS$_{2}$}

\title{\vspace{-1.0cm}\Huge\textbf{\textrm{\mytitle}}}  

\author{Siwen Zhao,$^{1,*}$ Jinqiang Huang,$^{2,3*}$ Valentin Cr\'epel,$^{4,*}$ Xingguang Wu,$^{5,6}$ Tongyao Zhang,$^{5,6}$ Hanwen Wang,$^{1}$ Xiangyan Han,$^{7}$ \\ Zhengyu Li,$^{8}$ Chuanying Xi,$^{8}$ Senyang Pan,$^{8}$ Zhaosheng Wang,$^{8}$ Kenji Watanabe,$^{9}$ Takashi Taniguchi,$^{10}$\\ Benjamin Sac\text{é}p\text{é},$^{11}$ Jing Zhang,$^{5,6}$ Ning Wang,$^{12,\dagger}$ Jianming Lu,$^{7,\dagger}$ Nicolas Regnault,$^{13,14,\dagger}$ Zheng Vitto Han$^{5,6,1,\dagger}$}

\date{} 
\begin{document}
\maketitle 
\vspace{-5mm}
\begin{center}
\begin{minipage}{1\textwidth}
\begin{center}
\par\textsuperscript{1} Liaoning Academy of Materials, Shenyang 110167, China
\par\textsuperscript{2} Center for Computational Quantum Physics, Flatiron Institute, New York, New York 10010, USA
\par\textsuperscript{3} Shenyang National Laboratory for Materials Science, Institute of Metal Research, Chinese Academy of Sciences, Shenyang 110016, China
\par\textsuperscript{4} School of Material Science and Engineering, University of Science and Technology of China, Anhui 230026, China
\par\textsuperscript{5} State Key Laboratory of Quantum Optics and Quantum Optics Devices, Institute of Optoelectronics, Shanxi University, Taiyuan 030006, China
\par\textsuperscript{6} Collaborative Innovation Center of Extreme Optics, Shanxi University, Taiyuan 030006, China
\par\textsuperscript{7} State Key Laboratory for Mesoscopic Physics, School of Physics, Peking University, Beijing 100871, China
\par\textsuperscript{8} Anhui Province Key Laboratory of Condensed Matter Physics at Extreme Conditions, High Magnetic Field Laboratory of the Chinese Academy of Science, Hefei 230031, Anhui, China. 
\par\textsuperscript{9} Research Center for Electronic and Optical Materials, National Institute for Materials Science, 1-1 Namiki, Tsukuba 305-0044, Japan
\par\textsuperscript{10} Research Center for Materials Nanoarchitectonics, National Institute for Materials Science,  1-1 Namiki, Tsukuba 305-0044, Japan
\par\textsuperscript{11} Universit$\Acute{e}$ Grenoble Alpes, CNRS, Grenoble INP, Institut N$\Acute{e}$el, 38000 Grenoble, France
\par\textsuperscript{12} Department of Physics and the Center for 1D/2D Quantum Materials, the Hong Kong University of Science and Technology, Hong Kong, China.
\par\textsuperscript{13} Laboratoire de Physique de L'Ecole normale sup$\acute{e}$rieure, ENS, Universit$\acute{e}$ PSL, CNRS, Sorbonne Universit$\acute{e}$, Universit$\acute{e}$ Paris-Diderot, Sorbonne Paris Cit$\acute{e}$, 75005 Paris, France
\par\textsuperscript{14} Department of Physics, Columbia University, New York, NY 10027, USA
\vspace{5mm}
\par{$\dagger$} Corresponding to: phwang@ust.hk, jmlu@pku.edu.cn, nicolas.regnault@phys.ens.fr, vitto.han@gmail.com
\par{$\star$} These authors contribute equally.
\vspace{5mm}
\end{center}
\end{minipage}
\end{center}

\newpage 
\clearpage

\setlength\parindent{13pt}
\noindent 
\section*{Abstract}
{\textbf{Semiconducting transition-metal dichalcogenides (TMDs) exhibit high mobility, strong spin–orbit coupling, and large effective masses, which simultaneously leads to a rich wealth of Landau quantizations and inherently strong electronic interactions. However, in spite of their extensively explored Landau levels (LL) structure, probing electron correlations in the fractionally filled LL regime has not been possible due to the difficulty of reaching the quantum limit. Here, we report evidence for fractional quantum Hall (FQH) states at filling fractions 4/5 and 2/5 in the lowest LL of bilayer MoS$_{2}$, manifested in fractionally quantized transverse conductance plateaus accompanied by longitudinal resistance minima. We further show that the observed FQH states sensitively depend on the dielectric and gate screening of the Coulomb interactions. Our findings establish a new FQH experimental platform which are a scarce resource: an intrinsic semiconducting high mobility electron gas, whose electronic interactions in the FQH regime are in principle tunable by Coulomb-screening engineering, and as such, could be the missing link between atomically thin graphene and semiconducting quantum wells.}}

\section*{Introduction}

Many-body effects driven by Coulomb interaction can lead to a wealth of incompressible gaps at partial fillings of Landau levels (LL), known as the fractional quantum Hall (FQH) states \cite{Laughlin_PRL_1983, Eisenstein_Science1990, Stormer_RMP1999, Halperin_RoPP_2021}. The FQH states are revealed through their edge conduction channels, whose transport exhibits a minimum of longitudinal resistance $\rho_\mathrm{xx}$ and a transverse conductance $\sigma _\mathrm{xy}$ quantized as $\nu e^{2}/h$, with $e$ and $h$ the elementary charge and the Planck constant, respectively. In the composite fermion theory \cite{Jain_PRL1989, Pan_PRL2003, Son_Review2018}, the main series of fractions is $\nu=p/(2kp + 1)$, with $k$ and $p$ integers, or its particle-hole conjugate partner $\nu=1-p/(2kp + 1)$.
FQH states are the very first known electronic system that hold promises for topological quantum computations, owing to their non-conventional exchange statistics in the non-Abelian 5/2 states observed in GaAs and ZnO semiconductors \cite{Moore_1991, Stern_AoP2008, Xie_NSR_2014}. Continuous theoretical efforts have also been devoted to the development of superconducting FQH hybrid systems to engineer parafermions, a generalized version of Majorana fermions, for quantum computing purposes with fractional topological orders \cite{Clarke_NC2013, Vaezi_PRX2014, Alicea_Review2016}.

For four decades, researchers have endeavored to search for solid state 2D electron gas (2DEG) systems that exhibit fractionally quantized $\sigma _\mathrm{xy}$ plateaus; a handful of them only, such as semiconductor quantum wells \cite{Tsui_PRL1982, NM_2010_ZnO, Falson_RoPP_2018} and graphene \cite{XuDu_Nature_2009, Cory_NatPhys_2011, Yacoby_2012, Patrick_Science, Yacoby_Science_2014, DGG_NC_2014}, being reported. The identification of new candidates for FQH remains a fundamental yet challenging pursuit with important consequences. For instance, the $SU$(4) symmetry breaking of LLs in graphene has lead to unprecedented collective states, including even denominator FQH states in the lowest LL and in higher LLs \cite{Leo_Science_2017, Young2017_BLG_Nature, Smet_NatPhys_2018, Young2018_BLG_NatPhys, JunZhu_PRX_2022}. Analogously, semiconducting TMDs display a rich variety of interacting phases due to the massive Dirac fermions character of their charge carriers and their strong spin-orbit coupling (SOC) \cite{NiuQian_PRL_2013, Ensslin_PRL_2018, Wang_NanoLett_2019}.

Recently, Landau levels in mono- or few-layered semiconducting TMDs have been intensively studied experimentally, manifesting giant effective $g$-factors \cite{Wang_PRL_2017, Mak_NN_2017, Tutuc_PRL_2017, Wang_NanoLett_2019}, non-conventional sequence of Hall plateaus \cite{Wang_NC_2016, Cory_NM_2018}, and LL crossings with possible quantum Hall antiferromagnetic phases \cite{Ensslin_PRR_2021}. Yet most of those studies are far above the quantum limit, $i.e.$ performed at filling fractions $\nu > 1$. Capacitance probes on high-quality monolayer WSe$_{2}$ has revealed a series of FQH gaps in the lowest and first excited LLs \cite{Cory_NN_2020}, suggesting that fractional quantizations are in principle available in TMDs for transport measurements, which is crucial for the future construction of quantum devices based on the FQH states. Still, fractional quantum Hall plateaus in TMDs, especially in the lowest LLs where electron interactions are the most pronounced, remains elusive. The major challenge lies in the concomitant lack of access to low carrier density close to the band edge, an Ohmic contact to the semiconducting channel at very low temperature, and an appropriate dielectric environment that favours the Coulomb interaction.

In this work, we report evidence for FQH at filling fraction 4/5 and 2/5 in bilayer MoS$_{2}$ via transport measurements under high magnetic fields up to 34 T and a base temperature of 300 mK. The technological key that enabled us to measure the FQH effect is new 2D-widowed bismuth contacts that provide Ohmic contacts to n-type MoSe$_{2}$ layers, down to the very low carrier density required to reach the quantum limit. Our calculations and measurements consistently suggest that, when subjected to a finite vertical electric field $E_\mathrm{z}$, the system behaves very much like a mono-layer MoS$_{2}$ whose first four (lowest and first excited) LLs in the conduction band are layer-valley locked, and the spin-degeneracy is fully lifted by the Zeeman energy. The specific LLs spectra of the MoS$_{2}$ bilayer, together with the asymmetric Ohmic contact to the high mobility bilayer semiconducting channel, thus provide a unique model to investigate the $e$-$e$ interactions in the FQH regime.

\begin{figure*}[t!]
\centering
\includegraphics[width=0.85\linewidth]{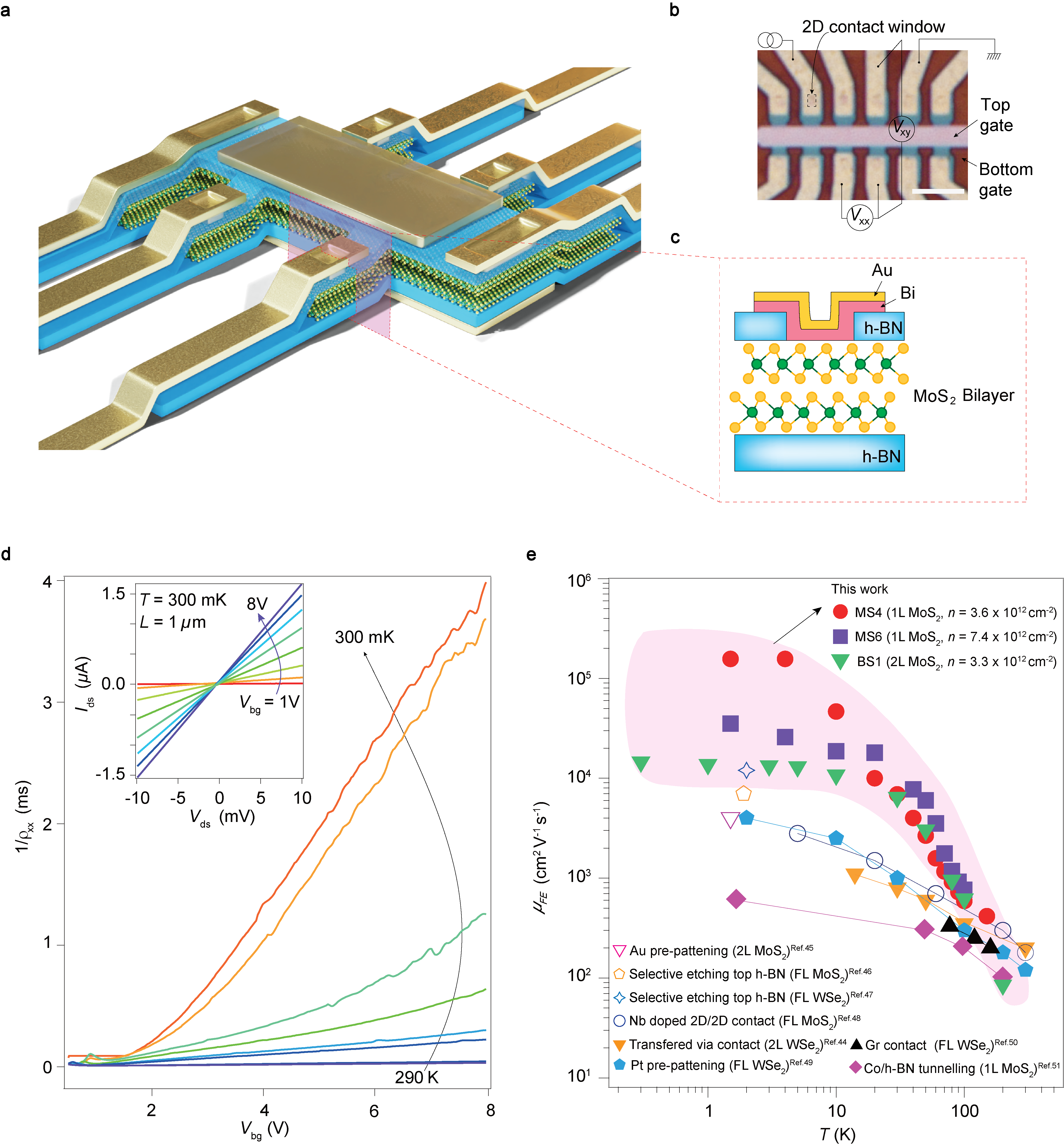}
\caption{\textbf{Electrical transport in bilayer MoS$_{2}$ in the single-particle regime.} (a) Schematic illustration of the fabricated device. (b) Optical image of a typical dual-gated bilayer MoS$_{2}$ device. One of the 2D contact windows is highlighted by black dashed lines. Scale is 5 $\mu$m. (c) Art view of the cross-section at the contact (indicated by the purple window in (a)) of the device under investigation. Bi 25 nm/Au 30 nm electrodes are deposited through the pre-patterned window in the top h-BN. (d) Field effect curves (longitudinal conductance $\rho_{\textrm{xx}}^{-1}$ $v.s.$ back gate voltage $V_{\textrm{bg}}$) of sample-BS1 measured at different temperatures from 290 K to 300 mK. Inset shows the two-terminal $I$-$V$ curves obtained at $T$ = 300 mK for a channel length of $L$ = 1 $\mu$m for different $V_{\textrm{bg}}$. (e) Summarization of state-of-the-art charge mobilities obtained in TMDs. Electron mobility in MoS$_{2}$ obtained in this work is the highest among TMDs transistors using different contact methods.}
\end{figure*}

\begin{figure*}[t!]
\centering
\includegraphics[width=0.85\linewidth]{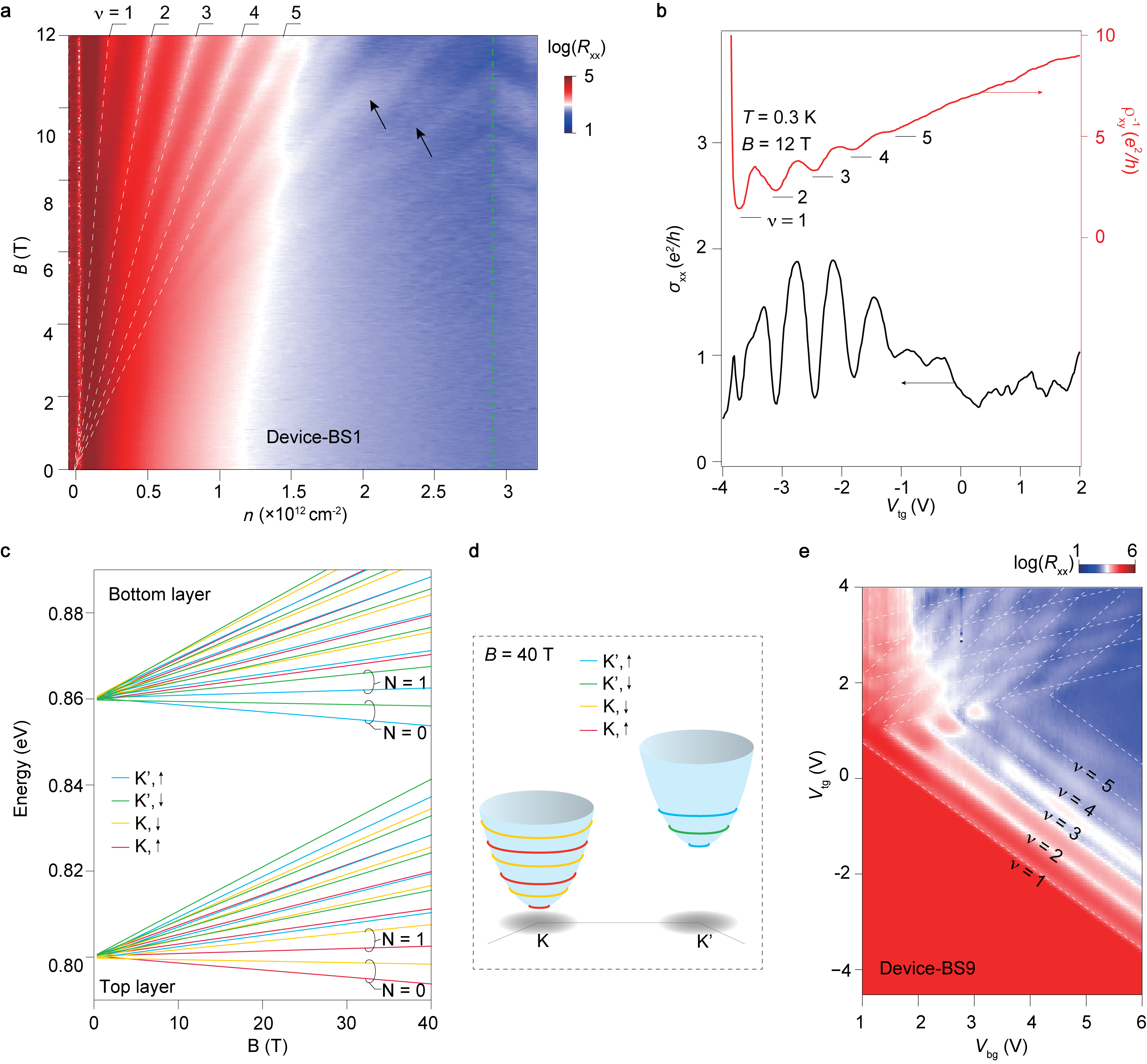}
\caption{\textbf{Electrical transport in bilayer MoS$_{2}$ in the single-particle regime.} (a) Landau fan of sample-BS1 measured at $T$ = 300 mK and $V_{\textrm{bg}}$ = 4.72 V, in the magnetic field range from 0 to 12 T. White dashed lines are guides to eyes, and their crossing point indicates the band edge. (b) Line profile of $\sigma_{\textrm{xx}}$= $\rho_{\textrm{xx}}$/($\rho_{\textrm{xx}}^{2}$ +$\rho_{\textrm{xy}}^{2}$) and transversal conductance $\rho_{\textrm{xy}}^{-1}$ at $B$ = 12 T, extracted from (a). (c) Calculated LLs as a function of energy and magnetic field (see details in Methods). (d) Cartoon illustration of the calculated first several LLs at the K and K' valley at $B$ = 40 T and $E_\mathrm{z}$ = 60 meV. For simplicity, only part of the Brillouin zone is shown. (e) Channel resistance  of an asymmetrically contacted bilayer MoS$_{2}$ sample device-BS9 as a function of $V_{\textrm{bg}}$ and $V_{\textrm{tg}}$. Dashed lines are guides to the eye.}
\end{figure*}

  
  \begin{figure*}[ht!]
  	\centering
  	\includegraphics[width=0.9\linewidth]{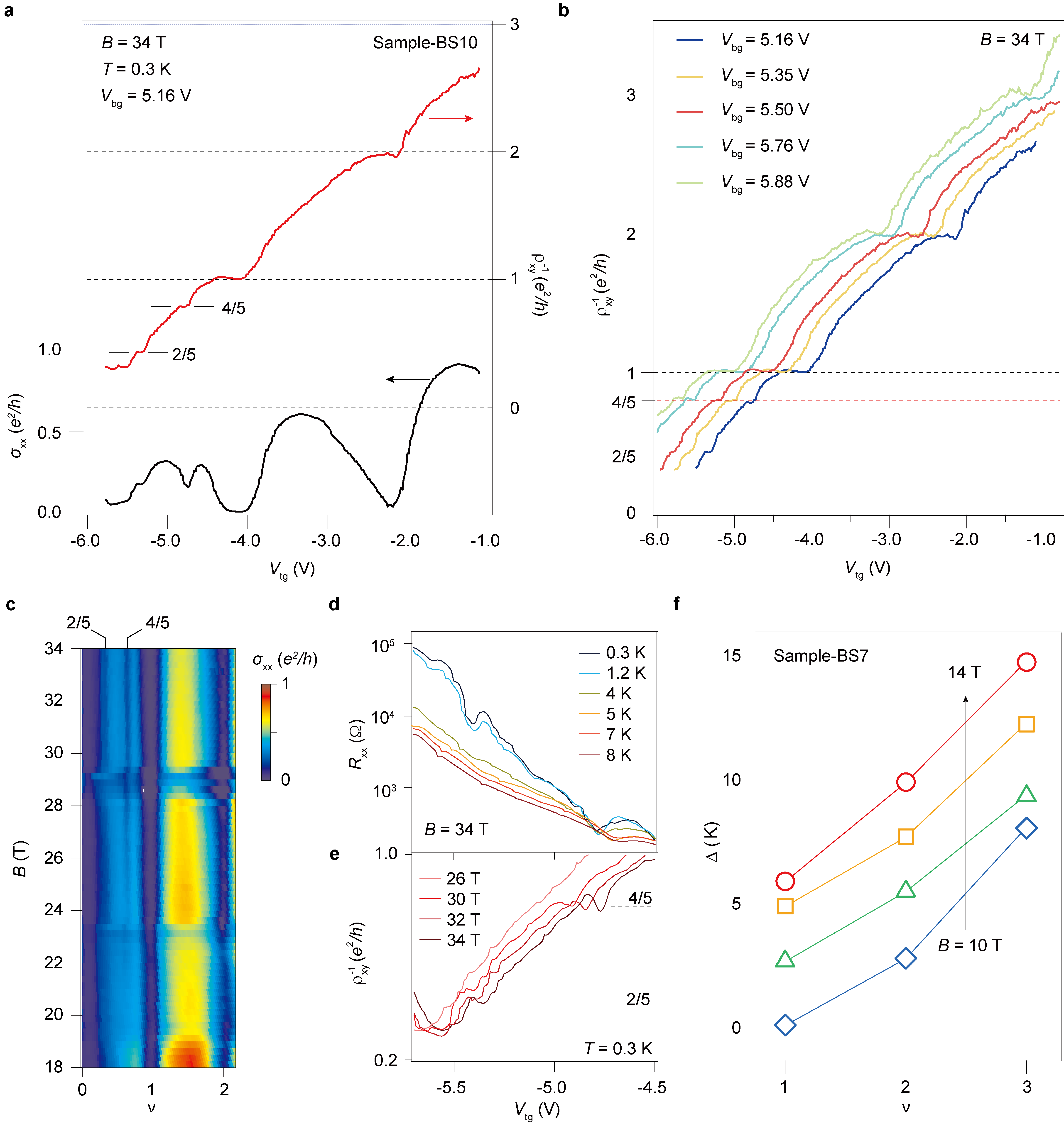}
  	\caption{\textbf{FQH states of bilayer MoS$_{2}$ in the lowest LL in conduction band.} (a) $\sigma_{\textrm{xx}}$= $\rho_{\textrm{xx}}$/($\rho_{\textrm{xx}}^{2}$ +$\rho_{\textrm{xy}}^{2}$), and $\rho_{\textrm{xy}}^{-1}$ as a function of $V_{\textrm{tg}}$. $V_{\textrm{bg}}$ is fixed at 5.16 V. (b) $\rho_{\textrm{xy}}^{-1}$ as a function of $V_{\textrm{tg}}$, measured for different $V_{\textrm{bg}}$. (c) Color map of longitudinal conductance measured in the filling fraction range of 0 to 2, and the magnetic field range of 18 to 34 T, at $T$ = 300 mK. (d) Raw data of $R_{\textrm{xx}}$ measured at different temperature. (e) Raw data of $\rho_{\textrm{xy}}^{-1}$ measured at different magnetic fields. $V_{\textrm{bg}}$ is fixed at 5.16 V in (d) and (e), and data in (a)-(e) are all obtained in bilayer MoS$_{2}$ sample-BS10. (f) Gaps of the first three LLs in the range of $B$ = 10 to 14 T in a typical bilayer MoS$_{2}$ device (sample-BS7). 
   }	
  	\label{fig:fig3}
  \end{figure*}

\section*{Results}

\noindent\textbf{Windowed Ohmic contact to n-type MoS$_{2}$ in the very-low carrier density limit.} \\ Although various methods, such as phase-engineered edge contact and the use of Pt or Au bottom electrodes \cite{Chhowalla_NM_2014, Wang_NC_2022, Ensslin_PRL_2018, DuanXiangfeng_Nature_2018}, have been attempted to establish electrical contact with semiconducting MoS$_{2}$ for quantum transport studies, the lowest LL approaching the quantum limit has not yet been possible through transport measurements. Recently, low-melting point metals like Bi and (01$\bar{1}$2)-faceted Sb have emerged as potential candidates for forming Ohmic contact to the MoS$_{2}$ channel \cite{XinranWang_Nature_2023}. However, these studies have been limited to relatively high-temperature regimes (above a few tens of Kelvin) thus far. Here, we adopt the bismuth contact method \cite{JKong_Nature_2021}, but modify it into a 2D-windowed convention.   

We first pre-pattern etched-through windows (with sizes of $\sim$ 1 $\times$ 1 $\mu$m$^{2}$) in few-layered hexagonal boron nitride (h-BN), which are further used as a top BN for encapsulating mono- or bi-layer MoS$_{2}$. The h-BN/MoS$_{2}$/h-BN sandwiches are fabricated in a N$_{2}$-filled glove box using the dry-transfer method \cite{Lei_Science_2013}. The work-flow of our sample fabrication process is illustrated in \textcolor{black}{Supplementary Figure 1}, with several typical samples shown in \textcolor{black}{Supplementary Figure 2}. Figure 1a shows an artistic view of the fabricated device. Thermal evaporation of Bi/Au (25 nm/ 30 nm) are carried out after a lithography that exposes the 2D-window opened in the top h-BN. As shown in the optical image of a typical device in Fig. 1b, the 2D-windowed contact (dashed box in Fig. 1b) is visible for each electrical leads. The device is equipped with top and bottom gates, and etch-shaped into Hall bars for electrical measurements (more details can be found in Methods). 

Figure 1c depicts a schematic of the windowed Bi contact on bilayer MoS$_{2}$, providing a cross-sectional view of the shaded section shown in Figure 1a. Clearly, the top surface of MoS$_{2}$ is contacted by Bi/Au in a 2D manner, which enables the formation of Ohmic contact to the conduction electrons in semiconducting MoS$_{2}$ down to base temperature of sub 1 K. As discussed in \textcolor{black}{Supplementary Figure 3}, by applying the transfer length method (TLM),  the contact resistivity in our devices is estimated to be 450 $\Omega \mu$m at $T$ = 1.5 K. In the typical dual gated samples, contact areas are doped by only the bottom gate, as illustrated in Fig. 1a-b and \textcolor{black}{Supplementary Figure 4a}. Taking sample-BS1 for example, contact resistance is estimated to be a few hundred Ohms by substracting 4-probe resistance from the 2-probe one at 30 mK, as shown in \textcolor{black}{Supplementary Figure 4b}. Such a contact resistance is sufficiently low in order to perform quantum transport measurements in the quantum Hall regime. Field effect curves (longitudinal conductance $\rho_{\textrm{xx}}^{-1}$ as a function of back gate voltage $V_{\textrm{bg}}$) recorded in a typical device BS-1 at different temperatures from 290 K to 300 mK are shown in Fig. 1d. It is seen that the device exhibits enhanced conductivity as the temperature decreases, and the two-terminal $I$-$V$ curves remain linear at the base temperature of 300 mK (inset in Fig. 1d), suggesting Ohmic contact to the n-type semiconductor.
Figure 1e compares the state-of-the-art mobilities obtained in typical TMDs transistors (n-type for MoS$_{2}$ and p-type for WSe$_{2}$, respectively) as a function of temperature using different contact methods \cite{jung2019transferred, pisoni2019absence, xu2017odd,wu2016even,chuang2016low,movva2015high,chuang2014high,cui2017low}. The electron mobility is estimated to be 10$^{4}$ \textendash 10$^{5}$ cm$^{2}$V$^{-1}$s$^{-1}$ in the typical MoS$_{2}$ devices in our work, benchmarking the best performance down to temperatures of sub 1 K. It is thus worth noting that our contact method may promote future applications of high electron mobiliry transistors (HEMT) devices using MoS$_{2}$ as a platform, especially at cryogenic temperatures.

\vspace{5mm}
\noindent\textbf{LLs of bilayer MoS$_{2}$.}\\ We now study the dual-gated mapping of the channel resistance of typical MoS$_{2}$ devices under finite perpendicular magnetic fields $B$. As shown in \textcolor{black}{Supplementary Figure 5}, the dual gate field-effect color maps at $B$ = 12 T and $T$ = 0.3 K are plotted into the parameter space of $n$ and $D$, with the total carrier density $n=(C_\mathrm{tg}V_\mathrm{tg}+C_\mathrm{bg}V_\mathrm{bg})/e-n_{0}$, and the applied average electric displacement field $D=(C_\mathrm{tg}V_\mathrm{tg}-C_\mathrm{bg}V_\mathrm{bg})/2\epsilon_{0} - D_{0}$, where $C_\mathrm{tg}$ and $C_\mathrm{bg}$ are the top and bottom gate capacitances per area, respectively. $V_\mathrm{tg}$ and $V_\mathrm{bg}$ are the top and bottom gate voltages, respectively. $n_{0}$ and $D_{0}$ are residual doping and residual displacement field, respectively. LLs close to the band edges are well developed for both mono-layer (\textcolor{black}{Supplementary Figure 5a}) and bilayer (\textcolor{black}{Supplementary Figure 5b}) MoS$_{2}$. Landau fan scanned at a fixed $D$ = -0.4 V/nm for the bilayer sample is given in \textcolor{black}{Supplementary Figure 5c}. It is seen that the lowest LL can be observed above $\sim$ 7 T, with the first 5 LLs indexed by fully degeneracy lifted integers $\nu$ from 1 to 5. Landau fan spectra measured in the same device but at a fixed $V_{\textrm{bg}}$ = 4.72 V is shown in Fig. 2a. Similar to that in \textcolor{black}{Supplementary Figure 5c}, the first 5 LLs are highlighted with dashed white lines, which extrapolate at $n$ $\sim$ 0 cm$^{-2}$, $i.e.$, the band edge. This extremely low carrier density available in our current system allows us to investigate the electronic states below the quantum limit of $\nu$ = 1. A line profile of the longitudinal and transverse conductance (defined as $\sigma_{\textrm{xx}}$= $\rho_{\textrm{xx}}$/($\rho_{\textrm{xx}}^{2}$ +$\rho_{\textrm{xy}}^{2}$) and $\rho_{\textrm{xy}}^{-1}$, with $\rho_{\textrm{xx}}$ and $\rho_{\textrm{xy}}$ the longitudinal and transverse resistivities; where a matrix inversion format is used for $\sigma_{\textrm{xx}}$ for better visibility due to the highly resistive $\rho_{\textrm{xx}}$ -- more details can be found in \textcolor{black}{Supplementary Figure 6}) are plotted in Fig. 2b, the quantization of Hall plateaus for the first 5 LLs can be seen with conductance minima observed in their corresponding $\sigma_{\textrm{xx}}$. Higher filling fractions become smeared out and level crossing seems to take place as can be seen in the doping range of $n$ > 1.5 $\times$ 10 $^{12}$ cm$^{-2}$, as indicated by solid black arrows in Fig. 2a. Notice that, the LL-crossing can be  complicated in multi-layered MoS$_{2}$, as reported previously \cite{NiuQian_PRL_2013, Wang_NanoLett_2019, Ensslin_PRR_2021}, but are nevertheless captured to a large extent by our calculations (\textcolor{black}{Supplementary Figure 7}). In addition, Fig. 2a displays another set of LL-like features (on the right side of the green dashed line), which is originated from the asymmetric electrical contact to the bilayer MoS$_{2}$, and will be discussed in the coming sections.

We now consider the LLs of bilayer MoS$_{2}$ devices. Their conduction band minima mainly consist of the $d_{z^2}$ electronic orbitals from both layers and are located at the $K$ and $K'$ corners of the Brillouin zone. At these points, the inter-layer hybridization carries non-zero orbital angular momentum and cannot couple the rotationally invariant orbitals of opposite layers.  
The layer index can therefore be taken as a good quantum number for electrons in the conduction band. 
Our calculations (\textcolor{black}{Supplementary Note 1}) suggest that, a finite vertical electric field $E_\mathrm{z}$ can split the layer index in the band structure of LLs, with the first four (lowest and first excited) LLs in the conduction band being layer-valley locked. A calculated band structure of LLs at layer polarization energy of $E_{\textrm{z}}=60$ meV is given in Fig. 2c, with more details of calculations displayed in the Suppl. Info. To visualize the calculated results, schematics of the first 9 LLs (calculated for $B$ = 40 T and $E_\mathrm{z}$ = 60 meV) in $K$ and $K'$ valley are illustrated in Fig. 2d, which largely differs from the $p$-doped scenario observed in monolayer WSe$_{2}$\cite{Cory_NN_2020}. 

One key feature of our study is that the Bi/Au electrodes are Ohmically contacted with only one top surface of the bilayer MoS$_{2}$, leading to an asymmetrically-contacted configuration. Unlike the monolayer case (\textcolor{black}{Supplementary Figure 5a}), this provides a specific manner to probe the LLs in each of the TMD layer, and enables the observation of LL crossing as shown in dual-gated scan in Fig. 2e and $D$-$n$ scan in \textcolor{black}{Supplementary Figure 5b}. When $E_{\textrm{z}}$ is biased to be sufficiently negative (for example, $D$ < -0.3 V/nm and $n$ < 5$\times$ 10 $^{12}$ cm$^{-2}$), the system effectively behaves as a mono-layer, with the $N$ = 0 and $N$ = 1 LLs being both in the same $K$-valley and on the top layer but carrying opposite spin polarization. In contrast, when $E_{\textrm{z}}$ is biased to be less negative or even positive, the lowest LLs generically feature a non-zero amplitude on the bottom layer. Due to the asymmetrically-contacted configuration, our measurements only probe a fraction of the total conductivity, leading to interference patterns that perfectly describe the anti-diagonal striped features (indicated by the white dashed lines in Fig. 2e), as also discussed in more details by comparing experimental and simulated results in \textcolor{black}{Supplementary Figure 8} and \textcolor{black}{Supplementary Figures 9-11}.

\bigskip
\noindent\textbf{FQH plateaus in high quality semiconducting bilayer MoS$_{2}$.}\\ In the following, by applying vertical magnetic field up to 34 T, at $T$ = 300 mK, we examine the QHE in bilayer MoS$_{2}$ down to the quantum limit of $\nu \leq$ 1. Figure 3a displays the $\sigma_{\textrm{xx}}$= $\rho_{\textrm{xx}}$/($\rho_{\textrm{xx}}^{2}$ +$\rho_{\textrm{xy}}^{2}$) and $\rho_{\textrm{xy}}^{-1}$ as a function of $V_{\textrm{tg}}$. Quantized plateaus at fractions of 4/5 and 2/5 are observed in the lowest LL, with a clear minima associated in the longitudinal conductance for the former and an emerging one for the latter. This makes semiconducting TMDs another platform to investigate, via electrical transport, FQH states after a few known 2DEG systems of quantum wells and graphene. By varying $V_{\textrm{tg}}$ at different values of fixed $V_{\textrm{bg}}$ ($i.e.$, different $D$), it is noticed that the FQH states have little dependence on the displacement field within the gate voltage range studied, as shown in Fig. 3b. This hints that, in agreement with earlier arguments, the observed FQH states are layer-polarized.

We then focus on the lowest LL with the magnetic field scanned from 34 T down to 18 T, as shown in Fig. 3c. Dips of $R_{\textrm{xx}}$ corresponding to fractionally quantized transverse conductance develop above $\sim$ 22 T at filling fractions 4/5 and 2/5. Raw data of $R_{\textrm{xx}}$ at different temperature, and $\rho_{\textrm{xy}}^{-1}$ at different magnetic fields, are plotted in Fig. 3d and 3e, respectively. As a function of temperature, we observe that the kink of $R_{\textrm{xx}}$ at filling fractions of $\nu$ = 2/5 rapidly vanishes near 2 K, and that at $\nu$ = 4/5 gradually disappears for $T$ up to 8 K. This means the FQH gaps are in the order of a few K at 34 T for the typical bilayer MoS$_{2}$ samples in this work. Meanwhile, the plateaus of $\rho_{\textrm{xy}}^{-1}$ (Fig. 3e) get smeared out below $\sim$ 26 T, same as that shown in Fig. 3c. It is worth mentioning that the FQH states with denominator 5 were reproduced in another typical bilayer MoS$_{2}$ sample (BS-6), as shown in \textcolor{black}{Supplementary Figure 12}. We further performed temperature dependence measurements of the quantum oscillations of the LLs (\textcolor{black}{Supplementary Figure 13}), and the gaps of the first three LLs in the range of $B$ = 10 to 14 T are extracted, as shown in Fig. 3f. Indeed, the gap size of fractionally-charged quasiparticles are smaller compared to those estimated at 34 T at integer filling (Fig. 3f).

\begin{figure}[t!]
\centering
\includegraphics[width=1\linewidth]{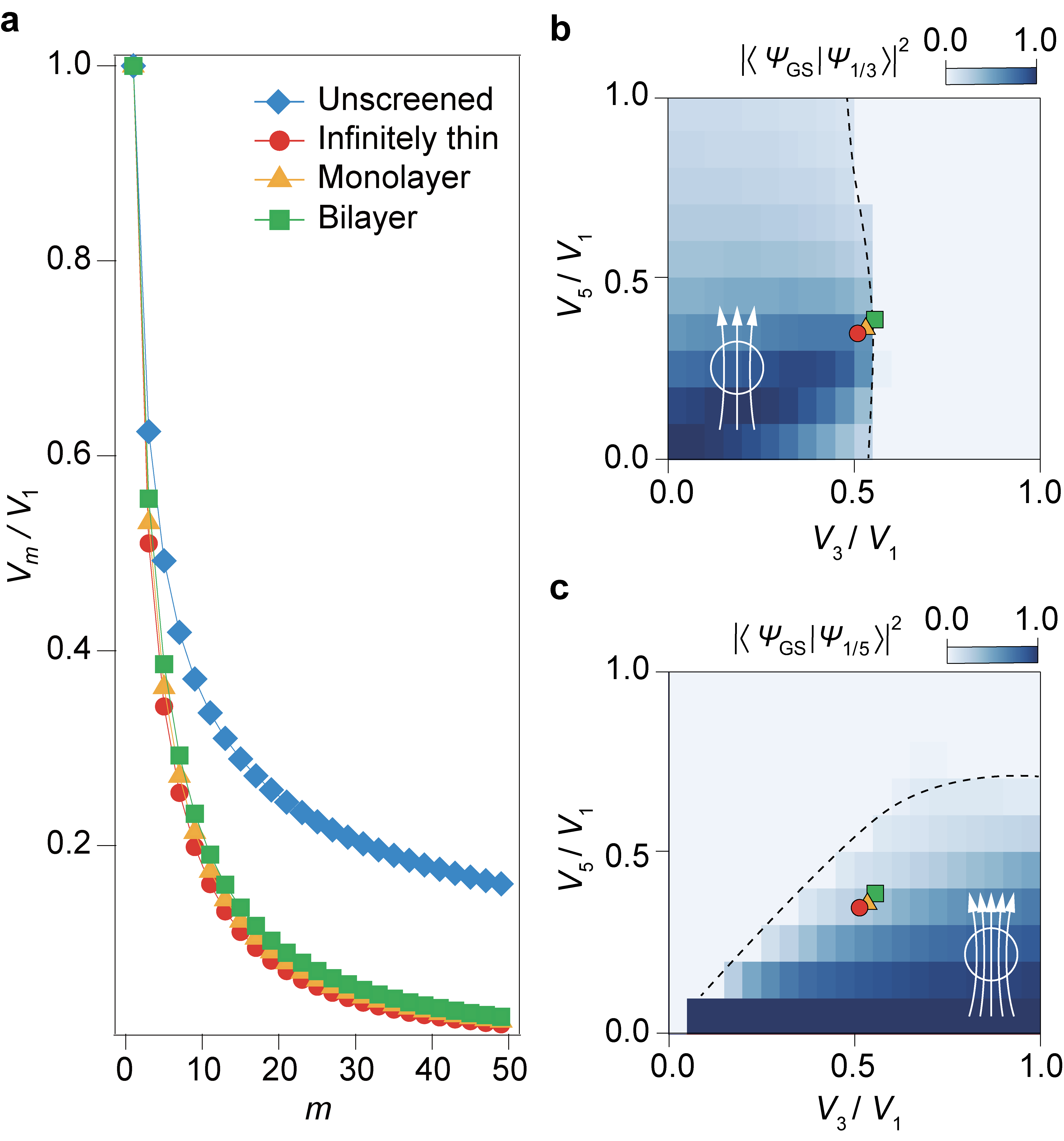}
\caption{\textbf{Numerical investigation of screening effects.} (a) Pseudopotential for a gate-screened Coulomb potential in an infinitely thin 2DEG, a monolayer and a bilayer MoS$_2$ at $B=22$ T and a gate-distance $d=35$ nm  (see Methods for details). We show the pseudopotentials of an unscreened Coulomb potential for comparison. 
(b) Square overlap between the exact diagonalization ground state $\ket{\Psi_{\rm GS}}$ and Laughlin's wavefunction $\ket{\Psi_{1/3}}$ numerically computed on a torus with $N=13$ electrons at filling fraction $1/3$ using three non-zero pseudopotentials $V_{m=1,3,5}$. Colored dots shows the values extracted from (a). 
(c) Same as (b) at filling fraction $1/5$ using $N=10$ particles. Cartoons in (b) and (c) illustrates 1/3 and 1/5 Laughlin's phase, respectively, with the black dashed lines highlighting the boundary between a Laughlin-like phase (non-zero overlaps, i.e., blue colored area) and a phase where the Laughlin state is absent (vanishing overlaps, i.e., white colored area).
}
\label{fig:fig4}
\end{figure}

The FQH state at $4/5$ in MoS$_{2}$ bilayers clearly revealed in our transport measurements falls in the universality class of the particle-hole (p-h) conjugate of Laughlin's $1/5$ state~\cite{Laughlin_PRL_1983}. This state hosts Abelian fractionally-charged quasiparticles, but differs from the fraction $1/3$ usually reported to have the largest gap in Laughlin's hierarchy. Our theoretical calculations show that the $N=0$ Landau level in MoS$_{2}$ bilayers is spanned by radial wavefunctions identical to those of standard semiconducting 2DEGs, and experiences a similar Zeeman effect. We thus expect similar interacting phases in both systems, including FQH states at $1/3$ (or its p-h conjugate $2/3$). Deviations from this behavior most likely come from non-universal characteristics of the samples, such as disorder, the effective dielectric environment induced by nearby gates, or the axial profile of the wavefunctions related to the finite sample thickness.

We have theoretically modeled the effects of gates and finite thickness as a screening of the Coulomb potential at long and short distances, respectively (see Methods). In the $N=0$ Landau level, this screened Coulomb potential is fully specified by its Haldane pseudopotentials $V_m$~\cite{haldane1983}, describing the highest energy that two particles with relative angular momentum $m$ can have in the lowest Landau level (only odd $m$ have an effect on fermionic states). In Fig.~\ref{fig:fig4}a, we plot these pseudopotentials at $B = 22$ T where the 4/5 fraction appears in Fig. 3c for an infinitely thin 2DEG, a monolayer and a bilayer MoS$_2$ using a realistic gate distance $d = 35$ nm. The presence of gates yields an exponential decay of $V_m$ at large $m$, much faster than the standard algebraic decay of the unscreened Coulomb potential. As a results, the few lowest pseudopotentials should already capture the physics of our gate-screened samples, and we restrict our attention to the three first ones $V_{m=1,3,5}$ from now on.

Fig.~\ref{fig:fig4}a also points out that the finite thickness of the sample increases $V_{m>1}$ compared to $V_1$. Due to their exact clustering properties, FQH states are sensitive to such changes in the short-range interaction physics. For instance, the ground state of a FQH system at filling $1/3$ has perfect overlap with the Laughlin state when $V_1>0$ and all $V_{m>1} = 0$, but this overlap decreases as $V_3$ and $V_5$ grow~\cite{regnault2017}, as shown in Fig. \ref{fig:fig4}b where we plot this overlap obtained for a finite torus using exact diagonalization (see Methods). For $V_3 \gtrsim 0.5 V_1$, this overlap vanishes, signaling a phase transition and a ground state that no longer belongs to the Laughlin universality class. The increase of $V_{m>1}/V_1$ in bilayer MoS$_2$, observed in Fig.~\ref{fig:fig4}a, drives the system to the edge of the phase that has non-zero overlap with Laughlin's 1/3 state, hinting at a potential absence of FQH state at $2/3$ in our experiments. Consistent with our experiments, the same calculation at filling 1/5 in Fig.~\ref{fig:fig4}c shows that the stability of fractions $1/5$, or its p-h conjugate $4/5$, barely changes as we go from an infinitely thin 2DEG to a realistic bilayer MoS$_2$. In conclusion, while all microscopic details of our samples are not necessarily captured by our calculations, they identify a clear trend that rationalizes the absence of the 2/3 FQH state due to internal screening of the Coulomb potential.  It is noticed that for the typical monolayer MoS$_{2}$ devices, Laughlin's 1/3 states are also absent, though the hints of FQH states in monolayer samples are exhibiting less well-defined quantization at filling 1/5 compared to the bilayer ones (see \textcolor{black}{Supplementary Figure 14}). The observed sensitivity of FQH states to the dielectric environment imposed by nearby gates and by the vertical profile of the heterostructure may allow to engineer a richer phenomenology of FQH states in TMD based heterostructures~\cite{papic2012numerical}. 

\bigskip

To conclude, we have achieved Ohmic contact to n-type bilayer MoS$_{2}$ with extremely low carrier density, and performed systematic transport studies under high magnetic fields up to 34 T and a base temperature of 300 mK. When subjected to a finite vertical electric field $E_\mathrm{z}$, the system behaves as a mono-layer MoS$_{2}$ whose first four LLs in the conduction band are layer-valley locked, and the spin-degeneracy is fully lifted by the Zeeman shift. The asymmetric Ohmic contact to the high mobility bilayer semiconducting channel allows us to obtain fractionally quantized Hall plateaus of 2/5 and 4/5 in the lowest LL above 26 T, which exhibit negligible tunability against $E_\mathrm{z}$, and are at the energy scale of sub 1 K, in agreement with theoretical predictions. Our observation of FQH plateaus in TMDs establishes, after the famed quantum well and graphene, a new material platform for the FQH states-based topological electronic systems. Our research paves the way for low-density transport experiments on TMDs heterostructures, which are necessary to harness the full potential of topological FQH phases, and the promises of their zero-magnetic field analog, the fractional Chern insulators, recently detected in twisted TMD homobilayers~\cite{Cai2023,Zeng2023, Xiaodong_FCI, Tingxin_FCI}.

\section*{Methods}

\textbf{Sample fabrication.} vdW few-layers of the h-BN/MoS$_{2}$/h-BN sandwich were obtained by mechanically exfoliating high quality bulk crystals. The vertical assembly of vdW layered compounds were fabricated using the dry-transfer method in a nitrogen-filled glove box. Hall bars of h-BN/MoS$_{2}$/h-BN devices were achieved by plasma etching. Electron beam lithography was done using a Zeiss Sigma 300 SEM with an Raith Elphy Quantum graphic writer. Top gates as well as contacting electrodes were fabricated with a thermal evaporator, with typical thicknesses of Bi/Au $\sim$ 25/30 nm.

\vspace{3mm}
\noindent\textbf{High magnetic field facilities.} A hybrid magnet with maximum of 34 T magnetic field and base temperature of 300 mK was used. The facility was equipped with water cooling system, and are maintained by the Steady High Magnetic Field Facilities, High Magnetic Field Laboratory, Chinese Academy of Science. 

\vspace{3mm}
\noindent\textbf{Electrical measurements.} The high precision of current measurements of the devices were measured using a Cascade M150 probe station at room temperature, with an Angilent B1500A Semiconductor Device Parameter Analyzer. Gate voltages on the as-prepared Hall bars were maintained by a Keithley 2400 source meter. During measurements, the TMD layer was fed with an AC $I_\mathrm{bias}$ of about 10 nA. Low frequency lock-in four-probe measurements were used throughout the transport measurements under high magnetic field and at low temperatures.

\vspace{3mm}
\noindent\textbf{Theoretical modellings.} The Landau level diagrams for monolayer and bilayer MoS$_2$ are obtained by applying minimal coupling on the $\vec k \cdot \vec p$ Hamiltonian for conduction electrons derived from \textit{ab-initio} simulations. The calculation follows from the formalism of Ref.~\cite{zubair2017quantum}, which is reviewed in the Suppl. Info. for completeness. We use values of the gap, effective mass, Ising spin-orbit coupling strength in the conduction and valence band, spin and valley $g$-factors, and uniform interlayer hybridization provided by the first principles calculations of Refs.~\cite{kormanyos2014spin,fang2015ab} and rounded off to the millielectronvolt scale. 

To describe the interacting phases of our system, we have used the Coulomb potential $V(q) = 2 \pi e^2 \tanh (q d)  / [\epsilon q (1 + n \alpha q \delta) ]$, which faithfully captures the long-distance (small $q$) and short-distance (large $q$) asymptotics respectively imposed by gate screening and the out-of-plane structure of MoS$_2$ multilayers. More precisely, this Coulomb potential reproduces ($i$) $V(q \to 0) = 2 \pi e^2 \tanh (q d) / (\epsilon q)$ expected in presence of gates symmetrically placed at a distance $d$ around the sample, with $\epsilon \simeq 5$ the relative dielectric constant of the surrounding h-BN; and ($ii$) a Rytova-Keldysh form $V(q \to \infty) = 2 \pi e^2  / [\epsilon q (1 + n \alpha q \delta) ]$ known to be relevant for TMDs~\cite{meckbach2018influence}, with $n$ the number of layers, $\delta=0.65$ nm the typical interlayer spacing, and $\alpha$ a material specific constant. Note that $n=0$ describes an infinitely thin 2DEG with no short-range screening. 

Interactions projected to the $N=0$ Landau level are accounted for using Haldane's pseudopotentials~\cite{girvin2002quantum}, defined on the plane as $V_m = (2\pi \ell_B^2)^{-1} \int_0^\infty \tilde{q} {\rm d} \tilde{q} V(\tilde{q} /\ell_B) L_m (\tilde{q}^2) e^{-\tilde{q}^2}$ with $\ell_B$ the magnetic length, $L_m$ the $m$-the Laguerre polynomial, and $V(q)$ the Coulomb potential. This should be understood as an angular Fourier transformation of the Coulomb potential after projection to the $N=0$ Landau level, such that $V_m$ energetically penalizes particles with relative angular momentum $m$. Note that fermions are only sensitive to the odd $m$ pseudopotentials. The FQH state at fraction 4/5 is observed in the magnetic field range $22-34$ T, for which the magnetic length lies between $\ell_B \simeq 4.5-6$ nm. Focusing on $B=22$ T as in Fig.~\ref{fig:fig4}a, the $\sim 35$ nm thick boron-nitride layers between our sample and gates yields $d/\ell_B \simeq 6$, while the coefficient $\alpha = 15/\epsilon$ taken from Ref.~\cite{van2018coulomb} leads to $\alpha \delta/\ell_B \simeq 0.4$. Both of these estimates have been rounded to their first significant digit to highlight their phenomenological nature. 

The phase diagrams in Fig. 4b-c were obtained by numerical exact diagonalization of the interacting problem on the torus with a square aspect ratio including the three first pseudo-potentials relevant for fermions $V_1$, $V_3$ and $V_5$. We set $V_1=1$ to fix the energy scale. The Hilbert space dimensions for $N=13$ particle at filling 1/3 and $N=10$ particles at filling 1/5 are 16 020 564 and 20 544 878, respectively, in the momentum sectors of their ground state.

\section*{\label{sec:level1}Data Availability}

The data that support the findings of this study are available at Zenodo, https://doi.org/10.5281/zenodo.8079020.

\section*{\label{sec:level2}Code Availability}

Exact diagonalization results were obtained using the open source and publicly available software \href{https://www.nick-ux.org/diagham}{DiagHam}.

\section*{\label{sec:level3}Acknowledgement}
 N.R. and V.C. are grateful to Z. Papic for fruitful discussions. This work is supported by the National Key R$\&$D Program of China (No. 2022YFA1203903, 2019YFA0307800, and 2021YFA1400100) and the National Natural Science Foundation of China (NSFC) (Grant Nos. 92265203, 12104462, 11974357, U1932151, 12204287, and 11974027). Z.V.H. acknowledges the support of the Fund for Shanxi “1331 Project” Key Subjects Construction. N.W. thanks the support from the Hong Kong Research Grants Council (Project No. AoE/P-701/20). J.L. thanks the Beijing Natural Science Foundation (Grant No. Z190011). K.W. and T.T. acknowledge support from the JSPS KAKENHI (Grant Numbers 20H00354, 21H05233 and 23H02052) and World Premier International Research Center Initiative (WPI), MEXT, Japan. B.S. and N.R. acknowledge support from the QuantERA II Programme that has received funding from the European Union’s Horizon 2020 research and innovation programme under Grant Agreement No 101017733. N.R. was also supported by the European Research Council (ERC) under the European Union’s Horizon 2020 research and innovation programme (grant agreement No. 101020833). The Flatiron Institute is a division of the Simons Foundation. B.S. was supported by the European Union's Horizon 2020 research and innovation program under the ERC grant \textit{SUPERGRAPH} No. 866365.

\section*{Author Contributions}
Z.V.H. and S.Z. conceived the experiment and supervised the overall project. S.Z., J.H., X.W., and X.H. carried out device fabrications; S.Z. and J.H. carried out all the electrical transport measurements; Z.L., C.X., S.P, and Z.W. assisted the measurements at high magnetic fields, with H.W. and T.Z. participated; K.W. and T.T. provided high quality h-BN bulk crystals. Z.V.H., S.Z., J.L., N.W., J.Z., B.S., V.C., and N.R. analysed the experimental data. V.C., and N.R. conducted the theoretical modelings and performed first principles calculations. The manuscript was written by Z.V.H. and V.C. with discussion and inputs from all authors.

\section*{Additional Information}
Competing interests: The authors declare no competing interests.

\bibliography{Ref_V1.bib}
\bibliographystyle{naturemag}

\end{document}